\documentclass[12pt]{iopart}
\usepackage[margin=1in]{geometry}
\usepackage{graphicx}
\usepackage[usenames, dvipsnames]{color}
\usepackage[colorlinks,pdfborder={0 0 0}, plainpages=false]{hyperref}
\usepackage{iopams, amsfonts, amsbsy, amssymb, amsopn, amstext}
\usepackage{xspace} % Sensible space treatment at end of simple macros
\usepackage{dcolumn}% Align table columns on decimal point
\usepackage{bm}% bold math
\usepackage{multirow} % Multiple rows in tables
\usepackage[raggedright]{subfigure}
\usepackage{ulem}

\definecolor{CiteColor}{rgb}{0, 0.5, 0} %
\hypersetup{citecolor=CiteColor} %
\definecolor{RefColor}{rgb}{0.55, 0, 0} %         
\hypersetup{linkcolor=RefColor} %

\definecolor {darkgreen}{rgb}{0.2, 0.7, 0.2}

\newcommand{\beq}{\begin{equation}}
\newcommand{\eeq}{\end{equation}}
\newcommand{\beqn}{\begin{eqnarray}}
\newcommand{\eeqn}{\end{eqnarray}}

\begin{document}

\title[Disk formation in the disruption of a neutron
  star by a nearly extremal black hole]{Massive disk formation in the
  tidal disruption of a neutron star by a nearly extremal black hole}

%% \author{Geoffrey Lovelace} \Cornell\GWPAC\Caltech
%% \author{Matthew D. Duez} \WSU
%% \author{Francois Foucart} \CITA
%% \author{Lawrence E. Kidder} \Cornell
%% \author{Mark A. Scheel} \Caltech
%% \author{B\'{e}la Szil\'{a}gyi} \Caltech

\author{Geoffrey Lovelace$^{1,2,3}$, Matthew D. Duez$^4$, 
Francois Foucart$^5$, Lawrence E. Kidder$^1$, 
Harald P. Pfeiffer$^{5,6}$, 
Mark A. Scheel$^3$, 
and B\'{e}la Szil\'{a}gyi$^3$}
\address{$^1$ Center for Radiophysics and Space
    Research, Cornell University, Ithaca, NY, 14853, USA}
\address{$^2$ Gravitational Wave Physics and
    Astronomy Center, California State University Fullerton,
    Fullerton, California 92831, USA}
\address{$^3$ Theoretical Astrophysics 350-17,
    California Institute of Technology, Pasadena, CA 91125, USA}
\address{$^4$ Department of Physics \&
    Astronomy, Washington State University, Pullman, Washington 99164,
    USA}
\address{$^5$ Canadian Institute for Theoretical
    Astrophysics, University of Toronto, Toronto, Ontario M5S 3H8,
    Canada} 
\address{$^6$ Fellow, Canadian Institute for Advanced Research}

\eads{%
  \mailto{glovelace@fullerton.edu}, %
  \mailto{m.duez@wsu.edu}, %
  \mailto{ffoucart@cita.utoronto.ca}, %
  \mailto{kidder@astro.cornell.edu}, %
  \mailto{pfeiffer@cita.utoronto.ca}, %
  \mailto{scheel@tapir.caltech.edu}, %
  \mailto{szilagyi@tapir.caltech.edu} %
}

\begin{abstract}
Black hole--neutron star (BHNS) 
binaries are important sources of gravitational 
waves for second-generation interferometers, and BHNS
mergers are also a proposed engine for short, hard gamma-ray bursts. 
The behavior of both the spacetime (and thus the 
emitted gravitational waves) and the neutron star 
matter in a BHNS merger depend strongly and nonlinearly 
on the black hole's spin. 
While there is a significant possibility that astrophysical black holes 
could have spins that are nearly extremal (i.e. near the theoretical 
maximum), to date fully relativistic simulations of BHNS binaries have 
included black-hole spins only up to $S/M^2$=0.9, which corresponds to 
the black hole having approximately half as much rotational energy as 
possible, given the black hole's mass. In this paper, we present a new 
simulation of a BHNS binary with a mass ratio $q=3$ and 
black-hole spin $S/M^2$=0.97, the highest 
simulated to date. We find that the black hole's large spin leads to the most 
massive accretion disk and the largest tidal tail outflow of any 
fully relativistic BHNS simulations to date, even exceeding the results implied 
by extrapolating results from simulations with lower black-hole spin. The 
disk appears to be remarkably stable.
We also find that the high black-hole spin persists until shortly before the 
time of merger; afterwards, both merger and accretion 
spin down the black hole.
\end{abstract}

\date{\today}


%\maketitle

\section{Introduction}
\label{sec:intro}

\subsection{Motivation}
Black hole--neutron star (BHNS) mergers are expected to be a major
source for the upcoming advanced interferometric gravitational-wave
detectors (Advanced LIGO, VIRGO, and KAGRA~\cite{LIGOScience,aVIRGO,
Somiya:2012}), with an expected event rate of the order of 
10 BHNS/yr~\cite{2010CQGra..27q3001A} (Note however that this rate is poorly constrained: 
pessimistic models predict as little as an event every 5 years, while with optimistic
assumptions, event rates of up to 300 BHNS/yr are theoretically possible).
If the mergers leave a massive accretion
disk around the black hole, they are also a promising setup for
short-duration gamma ray bursts (SGRB)~\cite{1991AcA....41..257P,
1992ApJ...395L..83N,Janka1999}, which
might be observed either directly or as an ``orphan'' SGRB afterglow. 
If matter is ejected during the merger (either from the tidal tail or
a disk wind), other signals are conceivable, including radio afterglows
and/or ``kilonovae''.  
%% (See~\cite{metzger:11} for a discussion of
%% electromagnetic counterparts.)
(See~\cite{metzger:11} for a discussion of
electromagnetic counterparts.)

It is to be expected that the parameters of BHNS systems will vary widely; 
the mass and spin
of the black hole, in particular, can be substantially
different from system to system.  Population synthesis 
studies~\cite{2008ApJ...682..474B,2010ApJ...715L.138B} and
mass measurements of stellar-mass black holes~\cite{2010ApJ...725.1918O} both favor
black-hole masses of 8--11$M_{\odot}$, but they are currently unable
to constrain the spin of the black hole, because this is set mostly
by the spin acquired at the hole's formation~\cite{2008ApJ...682..474B}. 
This spin at formation will depend on the progenitor of the black hole
and the complicated core collapse and accretion dynamics of the black
hole's birth; existing simulations can produce a wide range of
spins~\cite{2011ApJ...737....6S,2011PhRvL.106p1103O}.

The black-hole spin can strongly affect the outcome of the BHNS merger. 
Spin decreases the innermost stable circular orbit (ISCO) 
radius and specific angular momentum for prograde
circular orbits, and it increases them for retrograde orbits.  It is
expected that aligned black-hole spins will increase the mass of the
post-merger accretion disk.  Numerical simulations confirm this expectation,
finding that the disk mass increases dramatically with black-hole
spin~\cite{Etienne:2008re,Foucart:2010eq,Kyutoku:2011vz}.  
Indeed, for BHNS systems
with black-hole mass $\gtrsim 10 M_{\odot}$, only those with high
dimensionless black-hole spin appear capable of producing significant
post-merger accretion disks~\cite{Pannarale:2010vs,FoucartEtAl:2011,
Foucart2012}.  For a low mass system ($M_{\rm BH}\sim 4.5M_\odot$), 
our previous simulations
found that the mass remaining outside of the black hole after merger
increased from 5\% of the neutron-star mass
for a system with an initially nonrotating black hole to 39\% for a
system with a pre-merger black-hole dimensionless spin of
$\chi \equiv S/M^2=$0.9~\cite{Foucart:2010eq}.

It might appear that this previous study has already effectively covered
the range of possible black-hole spins and the consequent post-merger
behavior.  However, $\chi$ turns out to be a poor measure of how close a black
hole is to extremality.  For example, a black hole with $\chi=0.9$
has only about half the rotational energy of an extremal black hole
of the same mass.  More importantly for the accretion phase, the ISCO radius
shrinks significantly between $\chi=0.9$ and $\chi=1$ (from 2.3$M_{\rm BH}$
to $M_{\rm BH}$~\cite{1972ApJ...178..347B}).  Thus, increasing the spin in this
range can appreciably
change the mass of nuclear matter that is able to avoid prompt accretion. 
It can also increase the efficiency of energy extraction from the
accretion disk, $\eta \equiv L/\dot{M}c^2$, where $L$ is the luminosity 
of the disk, $\dot{M}$ is the accretion rate, and $c$ is the speed of light. 
Taking $\eta$ to be the
specific binding energy at the ISCO, as for a thin disk with no ISCO
shear, the efficiency for an extremal black hole is higher than that
of a $\chi=0.9$ black hole by a factor of 2.8.  Finally, we note that for
$\chi=0.9$, the ISCO is still outside the ergosphere; the ISCO and,
perhaps, significant amounts of disk matter, are only found inside
the ergoregion for $\chi>0.94$.  It is conceivable that new behavior
is possible for these ``extreme'' cases of accreting $\chi>0.94$ black holes.

Another important drawback of our previous study was that the disk was
evolved for only a short time after its formation.  A thick accretion disk
of nuclear matter with a mass roughly 10\% of the central black hole could
show interesting dynamics.  If the disk is sufficiently cool, self-gravity
could make the disk unstable to spiral density modes or fragmentation.  For
thin disks, stability is determined by the Toomre parameter
$Q_T=\kappa c_s/\pi G\Sigma$~\cite{1960AnAp...23..979S,1964ApJ...139.1217T}, 
where $c_s$ is the sound speed of the disk, 
$\kappa$ is the epicyclic frequency, 
$\Sigma$ is surface density, and $G$ is the gravitational constant.
Instabilities in
thick self-gravitating disks have been investigated numerically in both
Newtonian~\cite{1988MNRAS.231...97G,1992ApJ...388..451C,1994ApJ...420..247W,
1998MNRAS.297.1139M,2010arXiv1006.4624T}
and relativistic~\cite{2010PhRvL.104s1101M,rezzolla:10,
2011PhRvD..83d3007K,kiuchi:11,Korobkin:2012gj} physics. 
The instabilities catalogued in these studies are often very sensitive
to the angular momentum profile of the disk, especially the
runaway~\cite{1998A&A...331.1143A,2004MNRAS.349..841D}
and the Papaloizou-Pringle instabilities~\cite{1985MNRAS.213..799P}. 
Since the disk is thick and non-Keplerian, its angular momentum profile
can only be known by modeling its formation---the BHNS merger, in our case.

Extreme black-hole spin can also profoundly enhance the electromagnetic
signal from the post-merger system.  The ISCO binding energy is, as we
have mentioned, a strongly nonlinear function of $\chi$.  Models of
neutrino-dominated accretion flows have also shown that a high black-hole spin
leads to a higher neutrino luminosity, presumably making more energy available
for a gamma ray burst~\cite{Popham99,Setiawan2006,
2007ApJ...657..383C,2007PThPh.118..257S}.  There is also much more
rotational kinetic energy in the black hole that can be extracted via
the Blandford-Znajek process
(with the luminosity $L_{\rm BZ}\sim \chi^2$)~\cite{1977MNRAS.179..433B}.

\subsection{Summary and overview}

To test the limit of high black-hole spin, high post-merger disk-mass BHNS
mergers, we numerically model an extreme case.  For the case that we consider,
the pre-merger dimensionless black-hole spin is $\chi=0.97$, aligned with the orbital
angular momentum.  The mass ratio is 3:1, corresponding to a low black-hole 
mass system.  The neutron star has a low compaction 
$M_{\rm NS}/R_{\rm NS}=0.144$, corresponding to a neutron-star
radius $R_{\rm NS}\sim$14km for a $1.4M_{\odot}$
($R_{\rm NS}\sim$12km for a $1.2M_{\odot}$ neutron star).  This
system is not intended to represent a typical
black hole-neutron star binary but is a first step toward 
exploring the effects of nearly extremal black-hole spin in BHNS mergers.
The configuration we consider here illustrates the dynamics
possible at one extreme of the astrophysically allowable parameter
space, and with this spin and
compaction, we can compare to other results to isolate the
effects of extreme spin, since
the particular combination of mass ratio 3:1 and compaction 0.144
has been particularly well studied by numerical relativity for lower
spins, from $\chi=0$ to $\chi=0.9$. The effects of high black-hole 
spin could be even more important in BHNS with higher mass ratios or with 
more compact neutron stars, since 
systems with high enough mass ratio or low enough compaction would need 
nearly extremal black-hole spin for the neutron star to tidally disrupt
outside the horizon
(and thus for the emission of an electromagnetic 
counterpart to the gravitational-wave signal)~\cite{Foucart2012}. We intend to 
explore such configurations in future simulations.

We begin our evolution 5.5 orbits ($\approx 21 \mbox{ ms}$) 
before merger, when the binary
is still in quasicircular orbit.  We follow
the system until $\approx 27 \mbox{ ms}$ after merger; 
this is 13 orbital periods of
the maximum density region of the post-merger accretion disk, long
enough to let us clearly see the settling of the accretion disk to
near equilibrium.

As expected, the extreme spin leads to a very large post-merger
accretion disk.  Indeed, a full 60\% of the neutron-star matter
is able to avoid falling into the black hole during the initial
plunge and merger.  The vast majority of this matter settles into
a massive accretion disk.  We note that the early disk mass is
significantly larger (by about 50\%) than what was found even for
a $\chi=0.9$ system, indicating that the disk mass is very
sensitive to the black-hole spin when the spin is nearly extremal.

The matter evolution can be divided into phases, each described in
more detail below.  First, the neutron star is tidally disrupted,
leading to an outgoing tidal tail and an ingoing accretion stream. 
Second, the accretion stream circles the black hole and collides
with itself.
This shock rapidly heats the matter
and disrupts the accretion stream.  Third, the disk settles into
axisymmetric equilibrium, starting in the inner regions and proceeding
outward.  This requires redistribution of the orbital energy and angular
momentum to allow fluid elements to settle into circular orbit. 
In the outer disk, the fluid is nonaxisymmetric and has sub-equilibrium
average angular momentum.  The disk accretes into the black hole slowly
on a timescale of hundreds of milliseconds.

The rest of this paper is organized as follows. In Sec.~\ref{sec:methods}, 
we summarize the numerical methods we use to construct and evolve
initial data for a black hole and neutron star in a circular inspiral, with 
the black hole having nearly extremal spin. Then, in Sec.~\ref{sec:results}, 
we present and discuss the results of our simulation, including the 
emitted gravitational waves and the 
behavior of the black hole's apparent horizon (Sec.~\ref{sec:vacuum}),
the behavior of the accretion disk that forms following the neutron star's 
tidal disruption (Sec.~\ref{sec:accretiondisk}), and the unbound material
ejected by the merger (Sec.~\ref{sec:ejecta}). We briefly conclude in 
Sec.~\ref{sec:conclusion}.

\section{Methods}
\label{sec:methods}

\begin{figure}
\begin{center}\includegraphics[width=8.2cm]{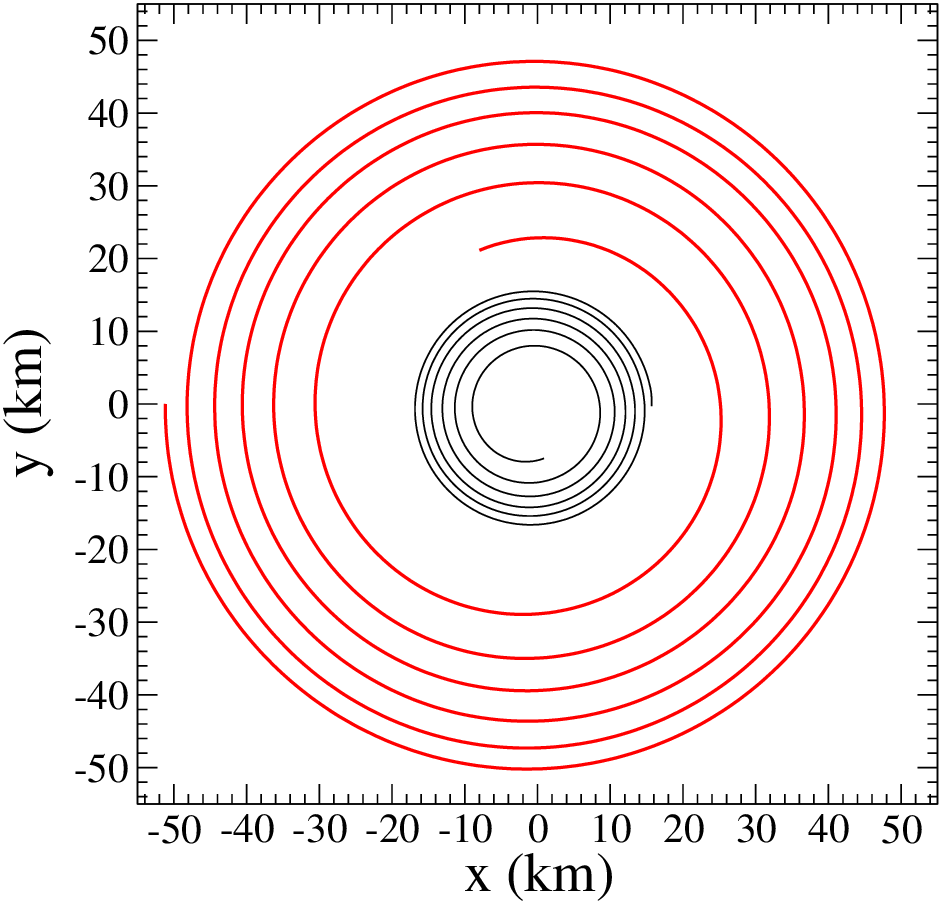}\end{center}
\caption{The trajectory of the apparent-horizon center (thin, black curve) and 
center of mass of the neutron star (thick, red curve) before tidal disruption 
(i.e., during the first $\approx 20$ ms 
after the start of the simulation). The orbital 
eccentricity in this case is $e = 0.005 \pm 0.001$.
\label{fig:TrajInspiral}}
\end{figure}

To construct initial data for a black hole--neutron star binary with
nearly extremal black-hole spin, we follow the methods
of Ref.~\cite{FoucartEtAl:2008}, which are in part motivated by the methods
for constructing binary-black-hole initial data with nearly extremal
spins in Ref.~\cite{Lovelace2008}.  We find that these methods, which
had only been tested for spins up to $\chi=0.9$ so far, can be
applied without any modification to generate initial configurations
with significantly higher black-hole spins --- although obtaining the
same accuracy as for lower spins requires the use of a finer numerical
grid close to the black hole. Here, we consider a black hole with
initial spin $\chi=0.970$. Because the initial conditions are
not perfectly in equilibrium, the spin slightly decreases during the
initial relaxation, falling to $a/M_{\rm BH}=0.967$ by time $t/M=100
M_{\odot}=1.5\mbox{ ms}$ after the beginning of the simulation 
(cf. Fig.~\ref{fig:BHMassAndSpinVsTimeWithMirr}). 
We use a polytrope
\begin{equation}
P = \kappa \rho^\Gamma + \bar{T} \rho,
\end{equation} where $P$ is pressure, $\rho$ is the rest-mass
density, and $\bar{T}$ is 
a fluid variable related to the physical temperature. As in 
Ref.~\cite{Foucart:2010eq}, we choose $\Gamma=2$ and choose $\kappa$ so that 
the compaction of the star is $M_{\rm NS}/R_{\rm NS} = 0.144$.
Finally, we reduce
the orbital eccentricity of the binary by iteratively solving for its
instantaneous angular velocity and inspiral
rate~\cite{Pfeiffer-Brown-etal:2007}.  The last iteration, which is
used throughout this paper, has eccentricity $e=0.005 \pm 0.001$. 
The orbital trajectories of the black hole and neutron star before 
tidal disruption 
(i.e., during the first $\approx 20$ ms of evolution after the start of the 
simulation, which corresponds to approximately 5.75 orbits), are shown 
in Fig.~\ref{fig:TrajInspiral}.

The coupled evolution of Einstein's equations of general relativity
and of the relativistic hydrodynamics equations is performed with the
SpEC code~\cite{SpECwebsite}, using our
two-grid method~\cite{Duez:2008rb}: Einstein's equations are evolved
in the generalized harmonic formalism~\cite{Lindblom2006}, using
pseudospectral methods and excision of the black hole interior, while
the neutron star fluid is evolved on a separate finite-difference grid
covering only the regions in which matter is present. The general
relativistic equations of hydrodynamics are evolved in conservative
form using a second-order finite volume scheme and high-order shock
capturing methods: the approximate Riemann problem is solved on cell
faces using the Weighted Essentially Non-Oscillatory (WENO5) reconstruction
algorithm~\cite{Liu1994200,Jiang1996202} and Harten, Lax, and van Leer (HLL)
fluxes~\cite{HLL}. Both sets of equations are evolved in time using
the third-order Runge-Kutta method with adaptive choice of the time
step.

To handle the black hole's high spin during the evolution, 
we find that we can apply the same techniques that have 
previously been applied to achieve fully relativistic simulations of 
merging black hole--black hole (BHBH) binaries with spins of 
$\chi=0.97$~\cite{Lovelace:2010ne,Lovelace:2011nu}. In particular, 
because we do not impose a boundary condition on the excision surface 
inside the black hole's apparent horizon, 
for the numerical evolution to be well-posed we must 
dynamically adjust the shape and velocity of the excision surface to 
prevent incoming characteristic fields~\cite{Hemberger:2012jz}. 
During the first $\approx10\mbox{ ms}$ 
of the evolution, we permitted the excision surface to fall well inside the 
apparent horizon, provided that there was no danger of incoming characteristic 
fields, but after noticing that this choice had the undesirable side effect 
of a noticeable increase in 
the measured Einstein constraint violation 
(cf. Fig.~\ref{fig:GRConstraintViolation}), for the remainder of 
the evolution we instead chose to 
adjust the excision surface so that it tracked the size of the apparent 
horizon. 

We simulated three resolutions through inspiral and tidal disruption, 
evolving $\approx 27 \mbox{ ms}$ (low, medium resolution) and $\approx 11 
\mbox{ ms}$ 
(high resolution) after merger. During the inspiral, these resolutions 
%correspond to approximately $56.9^3$, $64.5^3$, and $72.0^3$ 
correspond to approximately $57^3$, $65^3$, and $72^3$ 
spectral gridpoints, respectively, and $100^3$, $120^3$, $140^3$ 
finite-difference gridpoints, respectively. 
After the neutron star begins to tidally disrupt, we change our 
domain decomposition, using 
low, medium, and high resolutions with approximately
%$97.9^3$, $111.1^3$, and $124.4^3$ spectral gridpoints, respectively, and 
$98^3$, $111^3$, and $124^3$ spectral gridpoints, respectively, and 
$140^3$, $160^3$, $180^3$ finite-difference gridpoints, respectively. 
From then on, we use 
spectral adaptive mesh refinement to dynamically adjust the resolution 
of each spectral subdomain. Note that for all evolutions, since the 
black-hole spin is 
aligned with the orbital angular momentum, when solving the relativistic 
hydrodynamic equations we only evolve the region above the orbital plane, 
imposing a symmetry condition on the orbital plane and reducing the 
number of gridpoints in the direction normal to the orbital plane by 
a factor of 2.

We note that we do not find it necessary or advantageous to change 
to a damped harmonic gauge condition, as was done in the 
high-spin BHBH simulations 
in Refs.~\cite{Lovelace:2010ne,Lovelace:2011nu}. Instead, we use 
the same gauge conditions used in previous two-grid BHNS 
evolutions at the same mass ratio~\cite{Foucart:2010eq}. 

These simulations are computationally expensive. The low, 
medium, and high resolutions cost approximately $9.8 \times 10^4$, 
$1.4 \times 10^5$, and $1.5 \times 10^5$ CPU hours, corresponding to 
approximately 
$70$, $105$, and $116$ days of wallclock time. (Note that because we did not 
continue the high 
resolution simulation for as long after disruption, its expense is 
only slightly higher than that of the medium resolution.)

\section{Results}
\label{sec:results}

In this section, we present our main numerical results. First, in
Sec.~\ref{sec:vacuum}, we show results for the spacetime's evolution.
Then, we discuss the properties of the 
resulting accretion disk (Sec.~\ref{sec:accretiondisk}) and ejecta 
(Sec.~\ref{sec:ejecta}) after the black hole tidally disrupts the neutron star.

\subsection{Spacetime}
%\label{sec:gravitationalwaves}
\label{sec:vacuum}

\begin{figure}
\begin{center}\includegraphics[width=8.2cm]{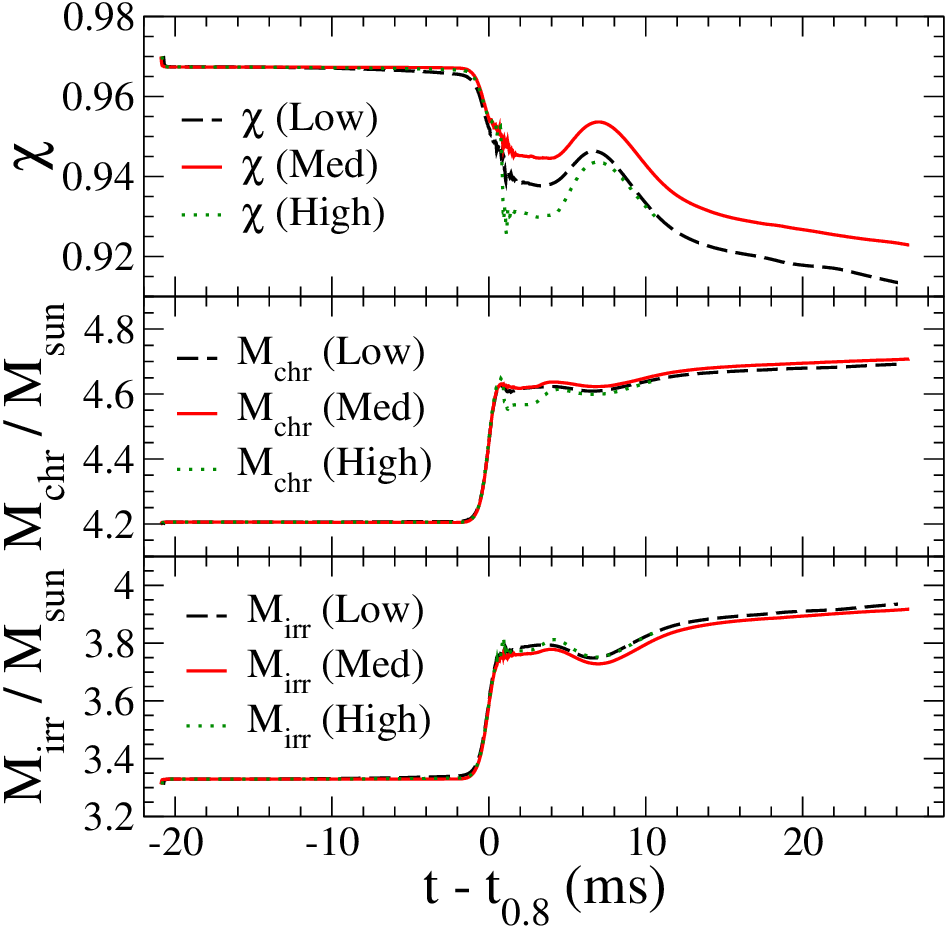}\end{center}
\caption{The evolution of the black-hole spin $\chi:=S/M_{\rm chr}^2$ 
(top panel), 
Christodoulou mass $M_{\rm chr}$ (middle panel), and irreducible mass 
$M_{\rm irr}$ (bottom panel) as 
a function of time (with times $t$ shown relative to time 
$t_{0.8}$, the time when the baryonic mass outside the hole has fallen to 
80\% of its initial 
value) for a black hole--neutron star merger with an initial black hole 
spin of magnitude $\chi=0.97$. 
\label{fig:BHMassAndSpinVsTimeWithMirr}}
\end{figure}

To characterize the behavior of the curved spacetime, 
we begin by examining the mass and spin of the black
hole as measured on the apparent
horizon. Figure~\ref{fig:BHMassAndSpinVsTimeWithMirr} shows, for each
resolution, the black-hole dimensionless spin $\chi := S / M_{\rm ch}^2$ 
(top panel), Christodoulou mass $M_{\rm ch} := \sqrt{M_{\rm irr}^2 + 
S^2 / \left(4 M_{\rm irr}^2\right)}$ (middle panel),
and irreducible mass $M_{\rm irr}$ (lower panel) 
as functions of time. Following the 
initial relaxation, the masses increase slightly 
(from $M_{\rm irr}=3.31$ and $M_{\rm ch}=4.20$ 
to $M_{\rm irr}=3.33$ and $M_{\rm ch}=4.21$) 
while the dimensionless spin 
decreases slightly (from $\chi=0.970$ to $\chi=0.967$). 
The masses and spins then remain constant in time to within our numerical 
accuracy until approximately $20\mbox{ ms}$ after the start of the 
simulation, when the neutron star begins to disrupt. During the next 
$\approx 1 \mbox{ ms}$,  
the masses sharply increase, while the dimensionless spin sharply decreases. 
The hole's masses then continue to slowly increase as its dimensionless 
spin slowly decreases.

\begin{figure}
\begin{center}\includegraphics[width=8.2cm]{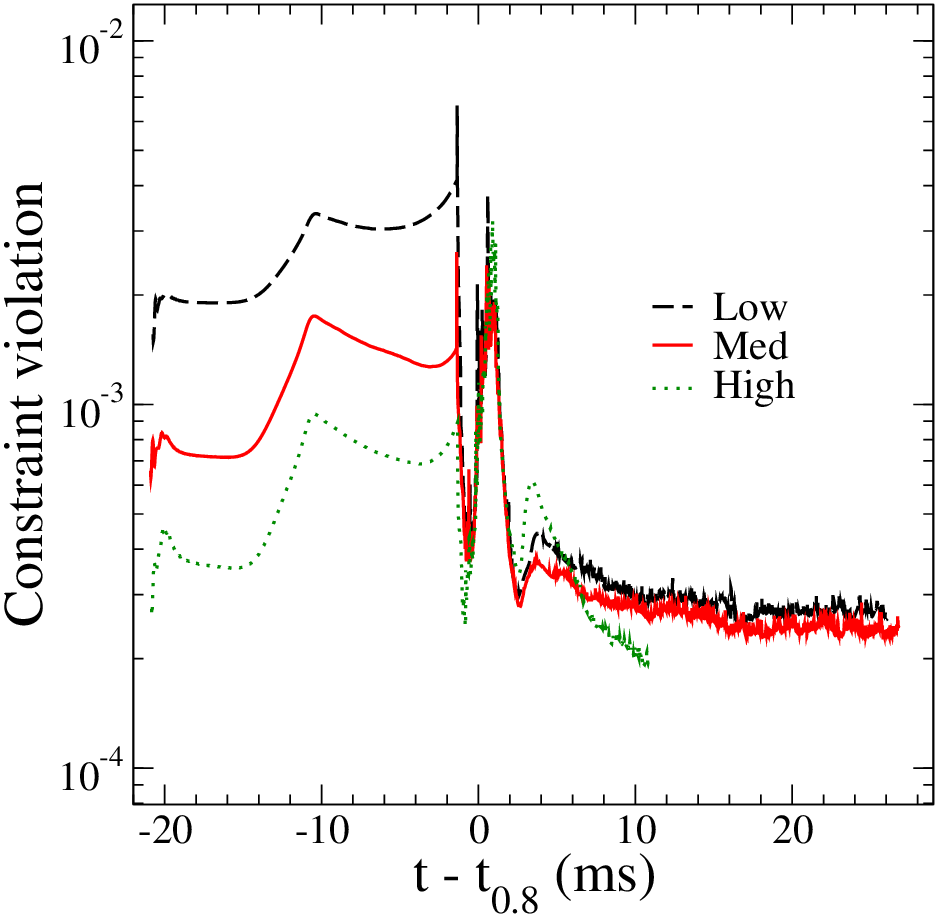}\end{center}
\caption{The normalized Einstein constraint violation  
(as defined in Eq.~(71) of Ref.~\cite{Lindblom2006}) 
as a function of time 
(with times $t$ shown relative to time 
$t_{0.8}$, the time when the baryonic mass outside the hole 
has fallen to 80\% of its initial 
value) for  
low (dashed, black), medium (solid, red), and high (dotted, green) resolutions.
In each simulation, the initial black-hole spin is aligned with the 
orbital angular momentum and has a magnitude of $\chi=0.97$. The constraints 
rise around time $t-t_{0.8}=-14\mbox{ ms}$ 
because the apparent horizon is moving 
farther from the excision surface. After modifying the simulation's algorithm 
so that the excision surface and apparent horizon would nearly coincide 
(at approximately $t-t_{0.8}=-10\mbox{ ms}$), 
the constraint violation drops until the neutron star begins to disrupt.
\label{fig:GRConstraintViolation}}
\end{figure}

While all three resolutions exhibit the same qualitative behavior, we were 
unable to demonstrate convergence of the black hole's mass and spin through 
the end of the simulation. Shortly after disruption begins, in the 
high-resolution simulation the black hole experiences an 
anomalously large decrease in the dimensionless spin; through 
experimentation, we found this decrease 
to be sensitive to the details of dynamic regridding (i.e., 
the algorithm used to resize the finite-difference grid as the region 
containing matter expands---cf. Sec. II B 1 of 
Ref.~\cite{Foucart:2010eq}). This could be a consequence of the 
extremely large amount of mass remaining in the disk.  Apparently,
there are certain special distributions of matter on our grid that,
upon a regrid,
result in unexpectedly large interpolation error near the 
excision surface and a corresponding
small but abrupt change in the evolution. 
Since our primary concern in this paper is the behavior of 
the disk and ejecta, and since this $O(\%)$ effect does not 
qualitatively change our results\footnote{
We are confident that this anomaly does not qualitatively affect our 
results or our conclusions, for the following reasons:
i) Cf.~the top panel of Fig.~\ref{fig:BaryonMass80}, which shows the 
result for all three $\chi=0.97$ resolutions: all three resolutions show 
the same qualitative behavior; ii) The $O(\%)$ 
anomalous behavior seen in the high-resolution black-hole quantities around 
time $t - t_{0.8} \approx 1\mbox{ ms}$ coincides with a period of the
evolution when the finite-difference grid rapidly, repeatedly regrids. 
The effect of these regrids can also be seen in small discontinuous
changes in disk properties such as the baryon rest mass.  
Disabling regridding did eliminate the anomalous behavior in the
high-resolution run.  Unfortunately, long evolutions cannot be done without
regriddng.  The frequent regridding behavior that seems to cause the problem
is not seen in the low and medium resolution runs.
iii) The low and medium resolutions, which don't display the frequent
regridding problem, show good quantitative agreement with each other.}, 
we have chosen to leave our efforts to resolve 
the black-hole spin after disruption 
(perhaps by improving our regridding method) for future work.

Figure~\ref{fig:GRConstraintViolation} further demonstrates our difficulty 
in obtaining convergence after tidal disruption. Before 
disruption (i.e. during the first 
$20\mbox{ ms}$ of the simulation), the normalized constraint violation appears 
to be convergent, as expected. The constraints do rise around 
$t-t_{0.8}\approx -14\mbox{ ms}$ 
as the excision surface falls far inside the apparent horizon, but this 
trend reverses after a modification 
of our method to keep the excision surface 
close to the apparent horizon\footnote{
Because 
of the high cost of our simulations (cf. Sec.~\ref{sec:methods}), we 
chose to adjust our algorithm during the simulations---at a time 
early enough that the constraint violation is still small---rather than 
to repeat the simulations from the beginning. 
We observe no significant effect corresponding to this adjustment 
in our results; therefore, we are 
confident that this adjustment does not affect our conclusions. 
However, had we continued to allow the excision surface to remain far 
inside the apparent horizon, constraint growth would eventually 
(before the time of merger) have caused the simulations to fail.
}. 
After interpolating to our 
higher-resolution domain (around $t-t_{0.8}\approx -2 \mbox{ ms}$), 
the constraints sharply fall at first, 
but then sharply spike during the merger. The subsequent bump in the 
constraints corresponds to the small bump in the dimensionless spin 
and the small drop in irreducible mass seen around $t-t_{0.8}
\approx 7 \mbox{ ms}$ in 
Fig.~\ref{fig:BHMassAndSpinVsTimeWithMirr}.

\begin{figure}
\begin{center}\includegraphics[width=11.2cm]{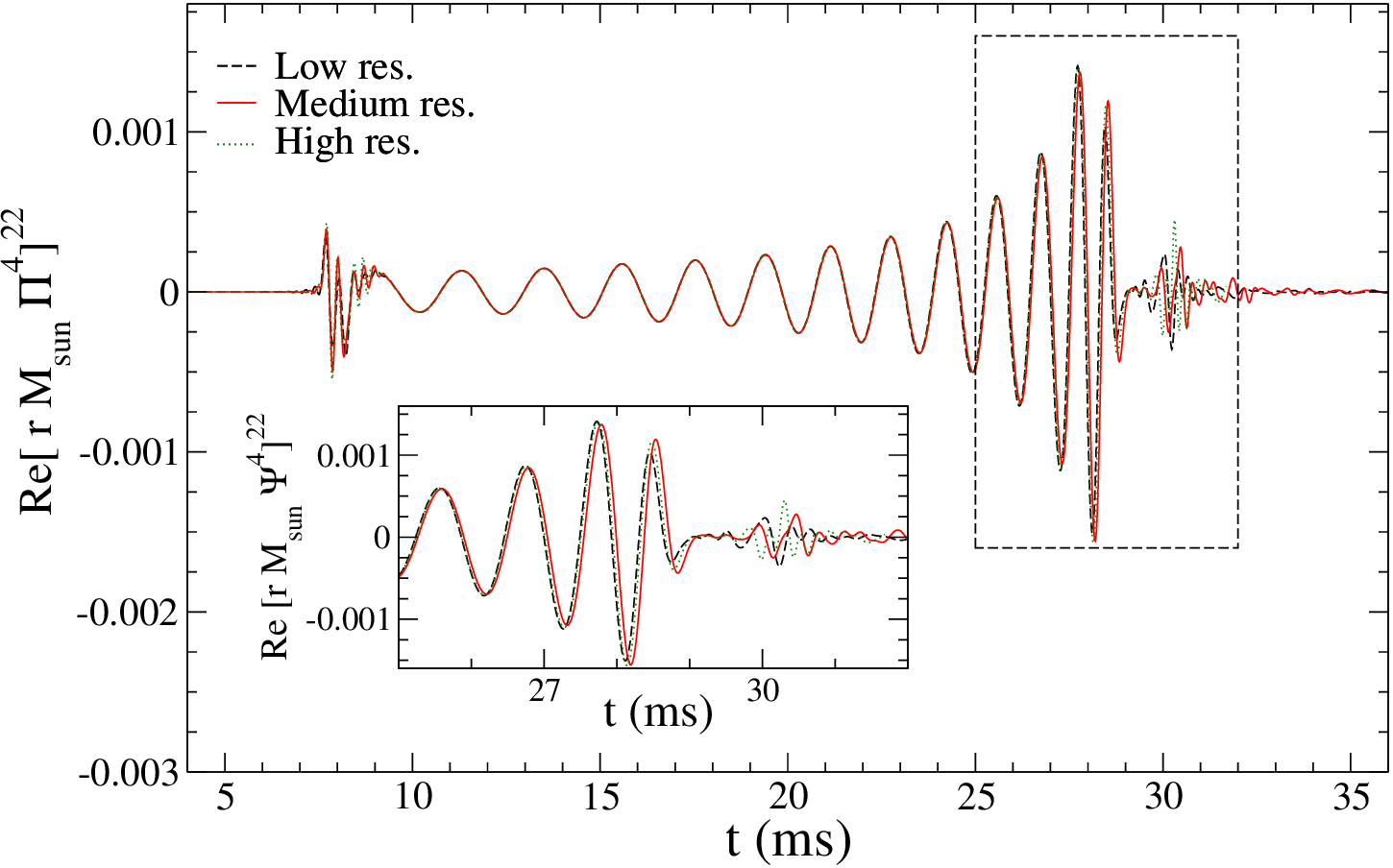}\end{center}
\caption{The real part of the dominant $\ell=m=2$ mode of the 
gravitational waveform $r M_{\rm sun} \Psi^4$, where $r$ is 
the radius of the sphere on which the gravitational waveform is 
extracted (here the outermost sphere, with a radius of 
$r=1540 M_\odot$), $M_{\rm sun} = 1 M_\odot$,  
and $\Psi^4$ is the Newman-Penrose 
scalar corresponding to outgoing gravitational waves. The inset zooms in on 
the part of the waveform enclosed by the dashed box.
\label{fig:GW}}
\end{figure}

We conclude this section by examining the dominant mode of the gravitational 
waveform (Fig.~\ref{fig:GW}). Because we plot $\Psi^4 = 
\left(d^2/dt^2\right)\left(h_+ - i h_\times\right)$ 
instead of the wave amplitude $h$, the 
spurious ``junk'' gravitational radiation emitted during the initial 
relaxation is clearly visible. Each resolution shows a small, secondary 
burst of 
gravitational waves after the primary wave has terminated. 
The phase of this
secondary burst cannot be resolved in the simulation: as is typical for numerical simulations
of BHNS or binary neutron star mergers, the phase error in the post-merger waveform
is on the order of a few radians. At all resolutions, the burst carries an energy
$E_{\rm burst} \sim 10^{-4}M_\odot c^2$ in a time $t_{\rm burst} \sim 0.75\,{\rm ms}$ and at
frequencies $f_{\rm burst}\sim 2-3\, \mbox{ kHz}$. There are no easy ways to
associate this burst in the gravitational wave signal extracted at large radius
with a specific feature of the forming accretion disk. Additionally, sharp features
due to shocks will only converge at first order with numerical resolution, and their
effects are likely to be poorly resolved. Nonetheless, an estimate from the
quadrupole formula shows that these numbers are consistent with emission from an
overdense region with excess mass $M_{\rm emitter}\sim 0.01-0.03 M_\odot$, 
orbiting in the inner region of the disk ($R\sim 30\,{\rm km}$). 
For the massive disk observed here,
the presence of such an asymmetry immediately after merger is quite natural.
We thus expect these features to be qualitatively correct, even though the detailed
properties of the waveform are not numerically resolved.

\subsection{Accretion disk}
\label{sec:accretiondisk}

\begin{figure}
\begin{center}\includegraphics[width=10.5cm]{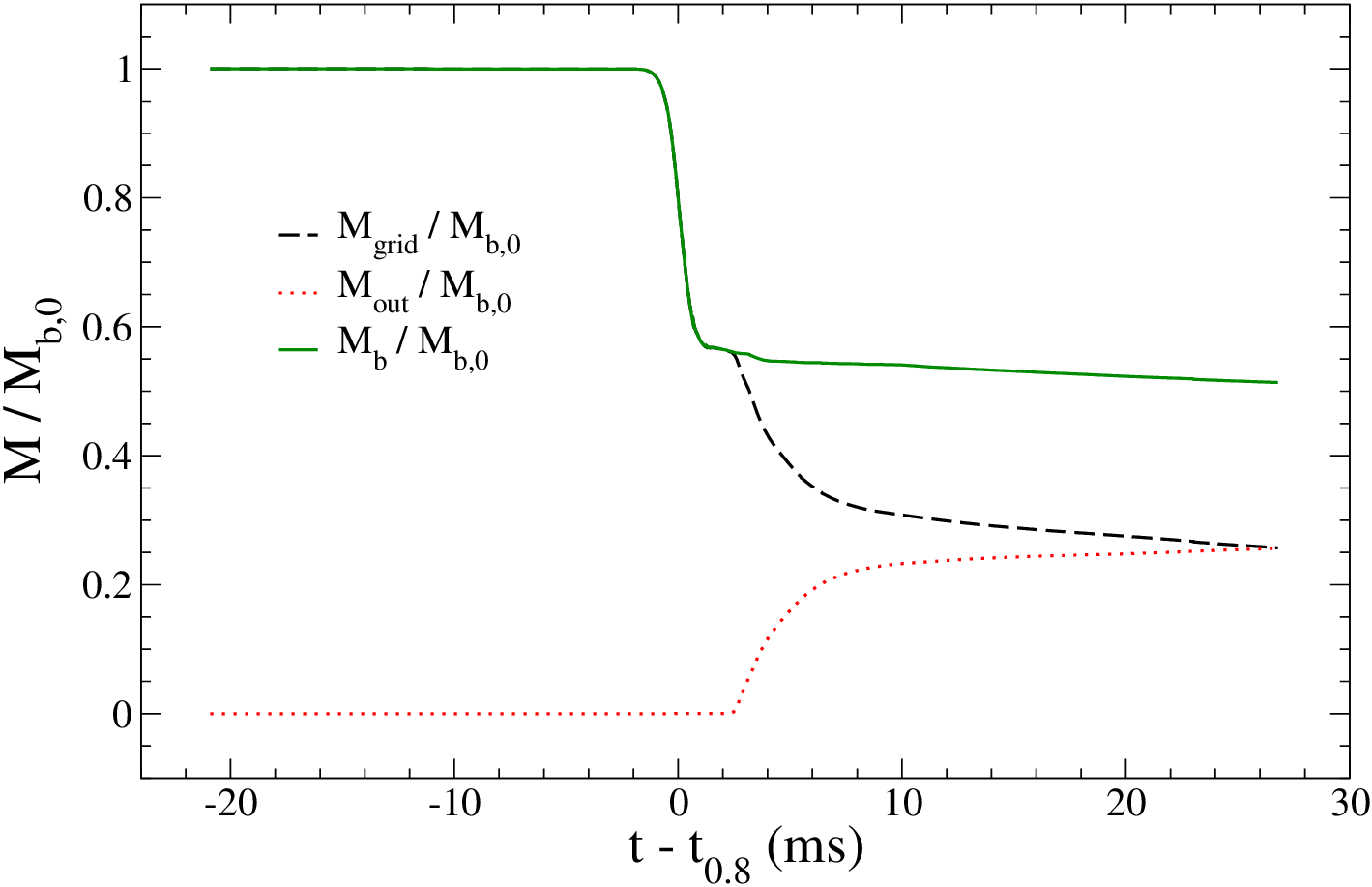}\end{center}
\caption{The evolution of the baryonic mass $M_{\rm b}$ 
as a function of time $t$ for the medium resolution black hole--neutron star 
merger with an initial aligned black-hole spin of magnitude $\chi=0.97$. 
The solid, green line is the total baryonic mass $M_{\rm b}$ 
outside the black hole,  
the dashed, black 
line is the total rest mass $M_{\rm grid}$ 
remaining on the finite-difference computational grid, and 
their difference $M_{\rm out}$ is the dotted, red line. All masses are 
normalized by the baryonic mass $M_{b,0}$ at the start of the simulation. 
Times are shown relative to time $t_{0.8}$, the time when the baryonic 
mass outside the hole has fallen to $80\%$ of its initial value.
\label{fig:Baryon80VsMassOnGrid}}
\end{figure}

After tidal disruption, the nuclear matter forms an outgoing tidal tail
and an ingoing accretion stream. 
At a time $t=t_{0.8} + 10\mbox{ ms}$ 
(10 ms after the time when 80\% of the initial baryonic mass remains 
outside the hole), $\approx 23\%$ of the initial baryonic mass has moved 
outward and left 
the finite-difference grid, 
and $\approx 31\%$ of the initial baryonic mass remains on the 
finite-difference grid (Fig.~\ref{fig:Baryon80VsMassOnGrid}).
Of the mass that has left, $\approx 17\%$ 
appears to be unbound and quickly leaves the grid (see Sec.~\ref{sec:ejecta}), 
while another
$\approx 6\%$ is weakly bound but extends far enough from the black hole to
leave the finite-difference grid.
This material would have fallen back onto the disk at
later times ($\sim 10^2$ ms) and its loss is one reason we limit our
simulations to the early to middle ($\le 30$ ms) post-merger evolution.

\begin{figure}
\begin{center}\includegraphics[width=8.2cm]{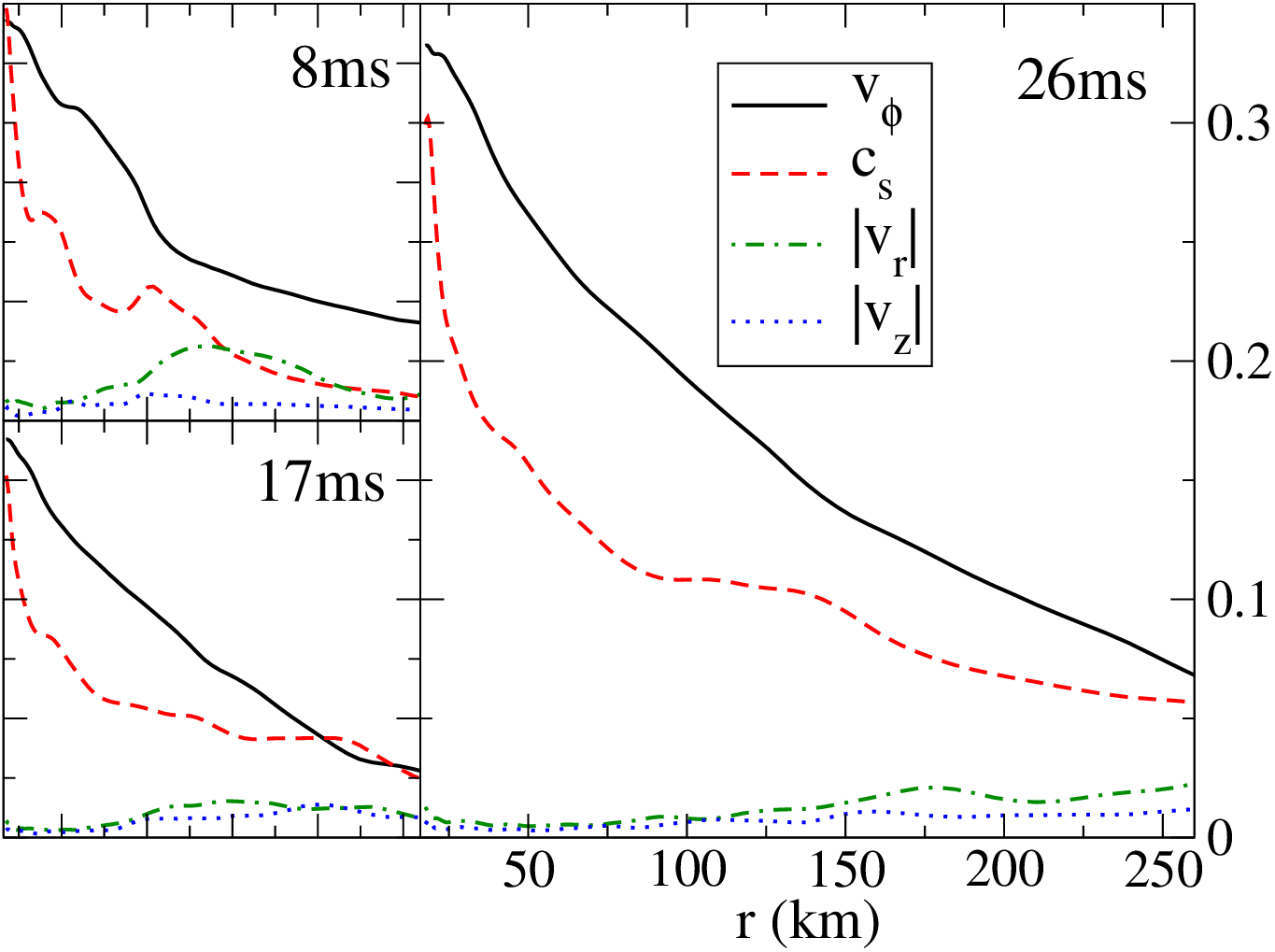}\end{center}
\caption{A break-up of the components of the fluid velocity,
shown at three times after merger, as a fraction of the speed of light.  
Also shown is the sound speed $c_s$. 
The extents on the horizontal and vertical axes are the same
in all three panels.  The radial velocity itself is negative (infall).}
\label{fig:velocity}
\end{figure}

During this tidal disruption phase, the matter remains cold and degenerate. 
The infalling stream of matter circles around the black hole and collides
with itself, producing an intense shock that travels through and disrupts
the incipient accretion flow.  Nearly all of the entropy generation in the
evolution occurs during this crucial millisecond, which could be labeled the
shock phase of the evolution.  Shock heating renders the matter completely
nondegenerate, and hot matter flows to the outer disk region ($r>100$km)
over the next $\sim 10$ms.  By the end of this time, a clear hierarchy
of velocities establishes itself through the disk out to 200km.  (See
Fig.~\ref{fig:velocity}.)  Inside this region, the azimuthal velocity
dominates and is roughly twice the sound speed throughout
($\Omega r\approx 2c_s$).  The orbital period sets the dynamical timescale;
we may define a ``settling radius'' $r_{\rm settle}(t)$ as the radius at which
gas has had time to complete exactly one orbital period since merger.  Since
orbits will not fully
circularize after one period, this radius provides only a rough sense of
which parts of the disk have had time to settle.  By the end of our evolution
(30\mbox{ ms} postmerger), $r_{\rm settle}=190$ km.  Outside $r_{\rm settle}$, the disk
does not have time to reach equilibrium. 
The radial velocity is a nonnegligible fraction of the azimuthal velocity
in the outer disk (see Fig.~\ref{fig:velocity}), but the radius at
which this infall becomes supersonic recedes with $r_{\rm settle}$. 

\begin{figure}
\begin{center}\includegraphics[width=10.2cm]{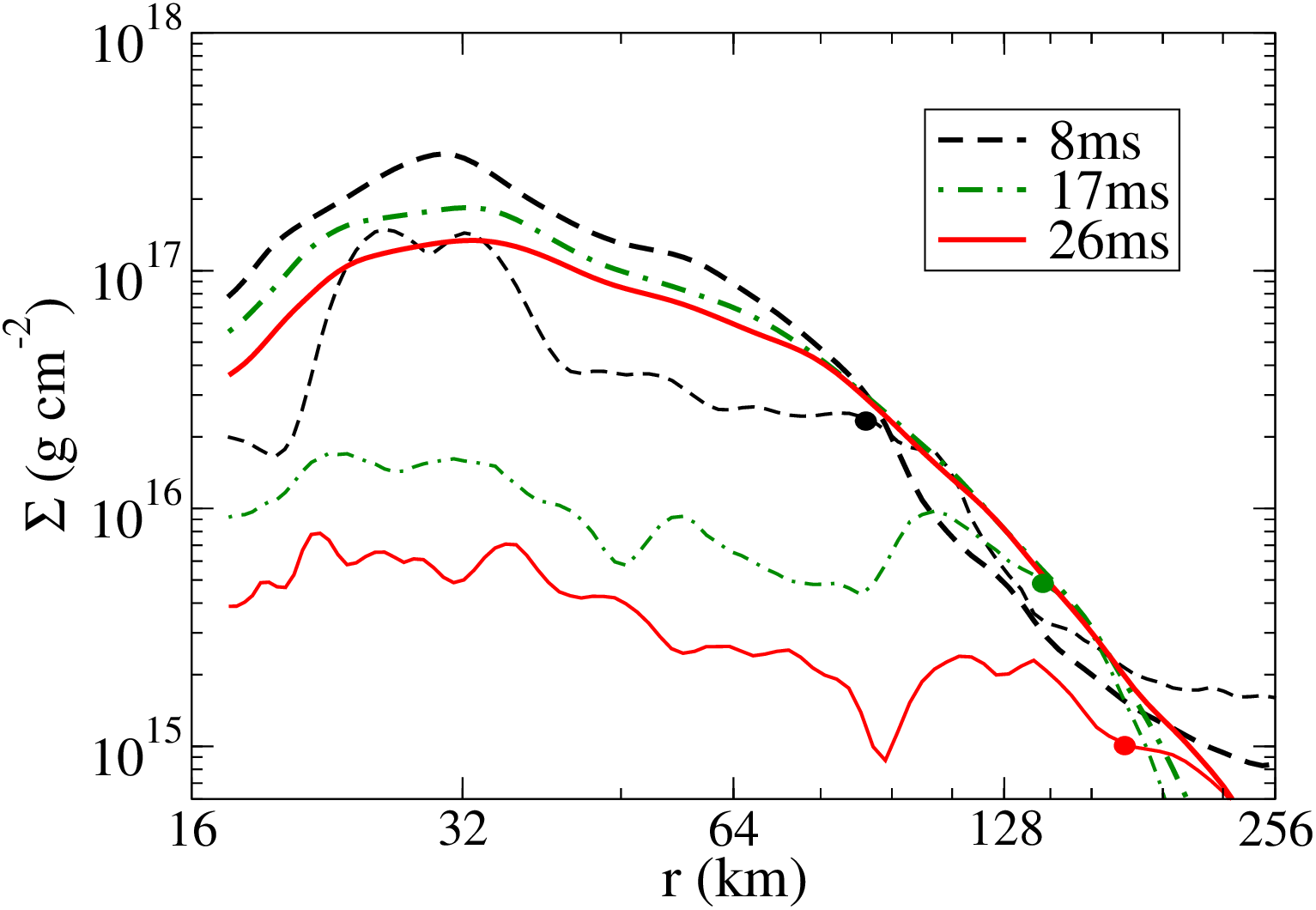}\end{center}
\caption{The density profile of the accretion disk, shown at
times $t=$8, 17, and 26ms after merger.  For each time, two
lines are shown.  The top line  is the vertically
integrated surface density $\Sigma$ at each radius, averaged over
all angles.  The bottom line is the rms deviation from the
average due to nonaxisymmetry.  A dot on the bottom line indicates
the location of $r_{\rm settle}$ at this time.  In this and all other radial
plots, the radius shown is constructed from the proper circumference
at the equator.}
\label{fig:density}
\end{figure}

In Figure~\ref{fig:density}, we show the surface density $\Sigma$ at three
times during
the settling phase.  The density peaks at a circumferential radius of around
30km; the location
of the peak remains fixed, while the overall density slowly decreases
everywhere as the mass is depleted by accretion.  Also shown in this plot
is the RMS deviation in $\Sigma$ at each radius due to deviations from
axisymmetry.  We see that the inner disk is quite axisymmetric, but that
deviations from axisymmetry reach order unity at around $r_{\rm settle}(t)$. 
A Fourier analysis of $\Sigma$ at late times shows
that the deviation from axisymmetry primarily subsists in trailing spiral
features, although localized twists in which a mode is leading over a small
radial range are occasionally seen in some modes.  The dominant mode in the
outer disk is $m=1$.  Only in the inner region can these modes be
regarded as linear perturbations of an equilibrium system; here they provide
a natural way to understand the extraction of angular momentum from the inner
disk which must occur if the low-angular-momentum outer disk (see below) is
to settle.  In addition to these smooth modes, equilibrium in
$r_{\rm settle}/2 < r < r_{\rm settle}$ is sometimes disturbed by weak,
localized sharp features, presumably weak shocks where gas on eccentric
orbits hits the more settled disk.  These features are most clearly seen
in the pressure force (see below).

In Figure~\ref{fig:equilibrium}, we plot the specific orbital energy
($E=u_t$) and angular momentum ($L=-u_{\phi}/u_t$) on the equator as a function
of radius.  For comparison, we plot the expected curves for circular
orbit geodesics in this spacetime metric.  The geodesic $E$ and $L$ 
are the expected orbital parameters for a stationary disk taking into
account the disk's self-gravity (which is included in the numerical metric)
but not pressure forces.  Except in the very inner disk, we see significant
deviations from geodesic orbits, as would be expected for a thick disk. 
Figure~\ref{fig:equilibrium} also includes $E$ and $L$ for equilibrium
circular orbits, i.e. stationary circular orbits taking into account
the pressure force
for the given fluid profile.  These agree with the actual $E$ and $L$
up to about 100km.  Beyond this, the energy curves continue to agree
(accounting for deviations due to localized shocks), but the angular
momentum profiles flatten more rapidly than the equilibrium curve
(although the latter seems to become somewhat noisy) and become
sub-equilibrium.  The gas at these radii was ejected (from the
tidal tail or the post-merger shock) into eccentric orbits, as seen
from the $L$ deficit and the nonzero radial velocity 
(cf. Fig.~\ref{fig:velocity}).  
Equivalently,
the nonaxisymmetric modes carry net negative angular momentum. 
Pressure support becomes especially important in the outer
disk, as can be seen from the deviation between geodesic and equilibrium
curves.  
We also have observed that the disk thickness, height $H$ divided by
radius $r$, increases with radius from about $H/r\approx 0.2$ in the inner
disk to about $H/r\approx 0.35$ in the outer disk.

\begin{figure}
\begin{center}\includegraphics[width=8.2cm]{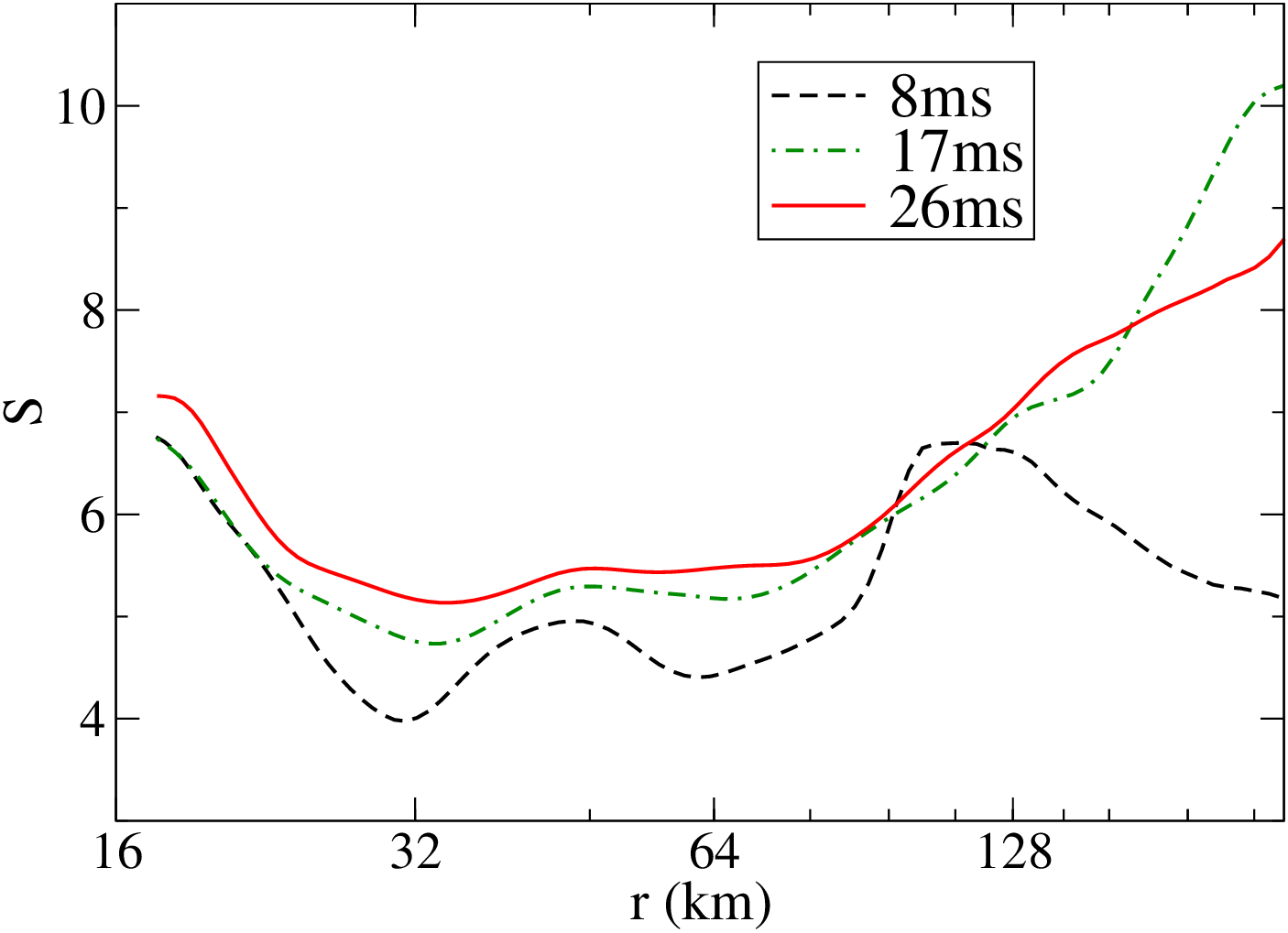}\end{center}
\caption{The entropy profile of the accretion disk, shown at
times $t=$8, 17, and 26ms after merger.  The entropy for our
$\Gamma$-law equation of state is $S=\log(\frac{P(\rho,T)}{P(\rho,0)})$.
For each radius, a
density-weighted average is carried out over the $\phi$ and
$z$ coordinates.
}
\label{fig:entropy}
\end{figure}

There are no clear signs of instability or turbulence in the disk,
a fact that itself should be explained.  The lack of obvious
global corotation instability of the kind found in some other
massive disk simulations is a natural consequence of the angular
momentum profile.  Instability is only expected for $L\propto r^n$,
$n<2-\sqrt{3}$~\cite{1985MNRAS.213..799P} with growth rate decreasing
rapidly as $n$ increases above zero, and our disk has $n\approx 0.3$
in the equilibrated region.  This region is also stable against the
effects of self-gravity, with Toomre parameter no lower than
$Q_T \approx 400$.  This stability is a consequence both of the shear
and the thermal pressure.  Radiative cooling may later remove some
of the disk's heat, but even removing all of it would leave $Q_T\sim 10$.
Shear and thermal energy also conspire to protect the disk from
convective instability.  In Newtonian physics, the Solberg-Hoiland
stability criteria are
\begin{eqnarray}
r^{-3}\nabla_rL^2 - (C_p\rho)^{-1}\nabla P\cdot\nabla S &>& 0 \\
\nabla_zP(\nabla_rL^2\nabla_zS - \nabla_zL^2\nabla_rS) &<& 0, 
\end{eqnarray}
where $\rho$ is the density, $S$ the entropy, $P$ the pressure,
and $C_P$ the heat capacity. 
The relativistic version of these equations is slightly more
complicated but basically similar~\cite{1975ApJ...197..745S}.  
Figure~\ref{fig:entropy} shows the average specific entropy as a
function of radius during the settling phase.  The entropy has a
minimum close to the density maximum, while the gas on the edges
is hotter.  Thus $\nabla P\cdot\nabla S <0$ except near a small
ring near $r=40$km, and this feature is stabilized by the strong
shear.  In fact, the epicyclic frequency is much greater than the
Brunt-V\"{a}is\"{a}l\"{a} frequency $N$ out to $r\approx 80$km. 
Beyond this, the
$S$ gradient steepens, the $L$ gradient flattens, and buoyancy
becomes an important restoring force.  Vertical convection is
also suppressed by a strong positive $\nabla_zS$.  The disk will
be unstable to the magnetorotational instability, since
$d\Omega^2/d\ln r + N^2<0$ throughout, but this
does not appear in our simulations, which do not include magnetic
fields.

The disk evolves slowly due to the settling of the outer disk and
accretion of gas in the inner disk (the two perhaps connected
by transfer of angular momentum).  The late-time accretion
rate is $\dot{M} \approx 2M_{\odot}$ s${}^{-1}$, giving a lifetime of
$\tau\equiv M/\dot{M} \sim 200$ms.  Before this time, we expect
magnetorotational effects to become important.  The most unstable
modes will grow on the dynamical timescale (ms), and then magnetic
turbulence will transport angular momentum.  Assuming this acts
like an $\alpha$ viscosity with $\alpha=$0.01--0.1,
the effective viscous timescale
will be of order $r^2/(\alpha c_sH)\sim$10--100ms for radius $r$, height
$H$, and sound speed $c_s$ characteristic of the central torus. 
Accretion increases the
spin and mass of the black hole, but since the accreted matter
has low angular momentum ($L\approx 10$, see Fig.~\ref{fig:equilibrium}),
the total effect on the dimensionless spin $\chi=S/M^2$ will be small 
even for accretion of most of the disk mass.  
(At late times in our evolution, $\chi$ is slowly decreasing.)

\begin{figure}
\begin{center}\includegraphics[width=8.2cm]{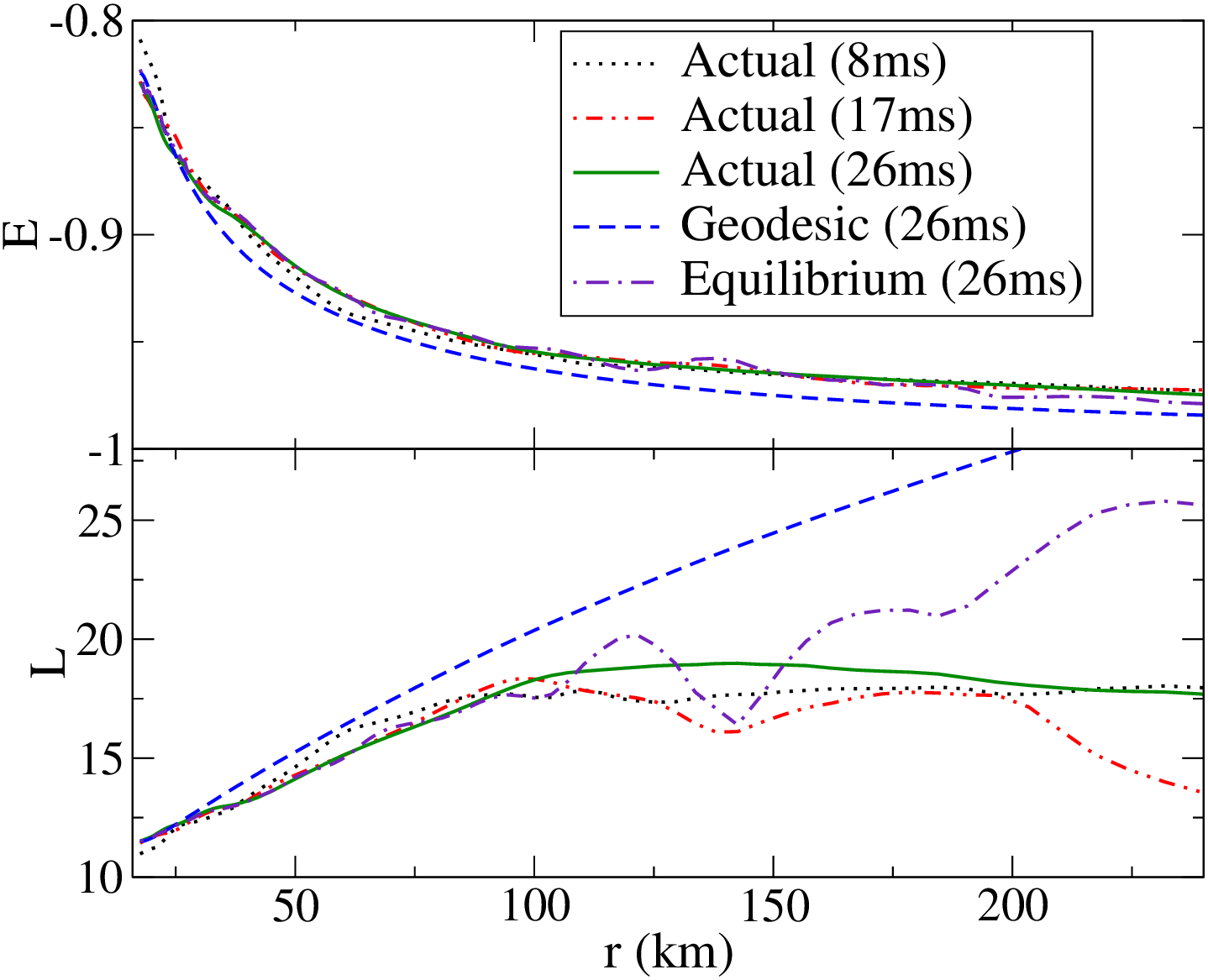}\end{center}
\caption{The specific orbital energy $E$ and specific angular
momentum $L$ at three times after merger.  For the final time,
we include two other curves for comparison.  First, we include
$E$ and $L$ for circular orbit geodesics in the given metric
(which will include the disk's self gravity).  Second, we
show $E$ and $L$ for equilibrium circular orbits given the
current metric and the pressure forces from the current pressure
and density profiles.  For the equilibrium energy, we take the
median in angle rather than the mean.  The mean would reveal spikes
at 140km and 190km due to local shock features.  The geodesic curves
for the earlier times
are nearly identical to those at late times, while the equilibrium
curves differ only at large radii.}
\label{fig:equilibrium}
\end{figure}

Our simulation does neglect several effects that could influence the
accretion disk's later evolution.  A magnetic field could produce a
jet, powered either by the accretion flow or the high black hole
spin.  It would certainly render our disk magnetorotationally
unstable.  Although we find that MRI turbulence is not needed for
strong early-time
accretion, it will still alter the disk evolution in important ways. 
Second, neutrino radiation will cool the gas and make it less neutron-rich. 
Assuming a simple ideal gas plus radiation pressure equation of state,
we infer that the temperature in the disk is of order 10MeV.  (In this
model, the disk is gas pressure-dominated everywhere except the inner
edge.)  Using a simple diffusion approximation, one can estimate a
luminosity $L_{\nu}\sim 10^{52}$erg s${}^{-1}$, which could deplete
the disk's thermal energy on a timescale of $\sim 10^2$ms.  More likely,
the cooling will balance the dissipative heating introduced by magnetic
turbulence.  Neutrino radiation could itself induce unstable entropy
or composition gradients and drive convection.  Finally,
our grid limitations have forced us to allow gravitationally bounded, 
outward-moving material to leave the finite-difference grid.  
This matter should eventually fall back
onto the disk, perturbing and perhaps shocking it.  Tracking
the fallback matter will be a computational challenge for future simulations
that wish to evolve disks like this one further in time
than was done here.

%\begin{figure}
%\caption{Fraction of the neutron star mass remaining outside of the
%  black hole after merger in numerical simulations at $q=3$, $M_{\rm
%    NS}/R_{\rm NS}=0.144$. Error bars are shown for the simulations
%  performed using the SpEC code. The point at $\chi=0.75$ is taken
%from Etienne et al.~\cite{Etienne:2008re}.
%  The solid line correspond to the
%  theoretical predictions derived in Ref.~\cite{Foucart2012}.
%\francois{Modified the error bars for $\chi=0.9,0.97$, using $15\%$ relative
%error. Is that right for $\chi=0.97$? Also removed the erroneous error
%bar from the UIUC results. FF}
%}
%\label{fig:mass_predict}
%\begin{center}\includegraphics[width=8.2cm]{MassPred.eps}\end{center}
%\end{figure}

\begin{figure}
\begin{center}\includegraphics[width=8.2cm]{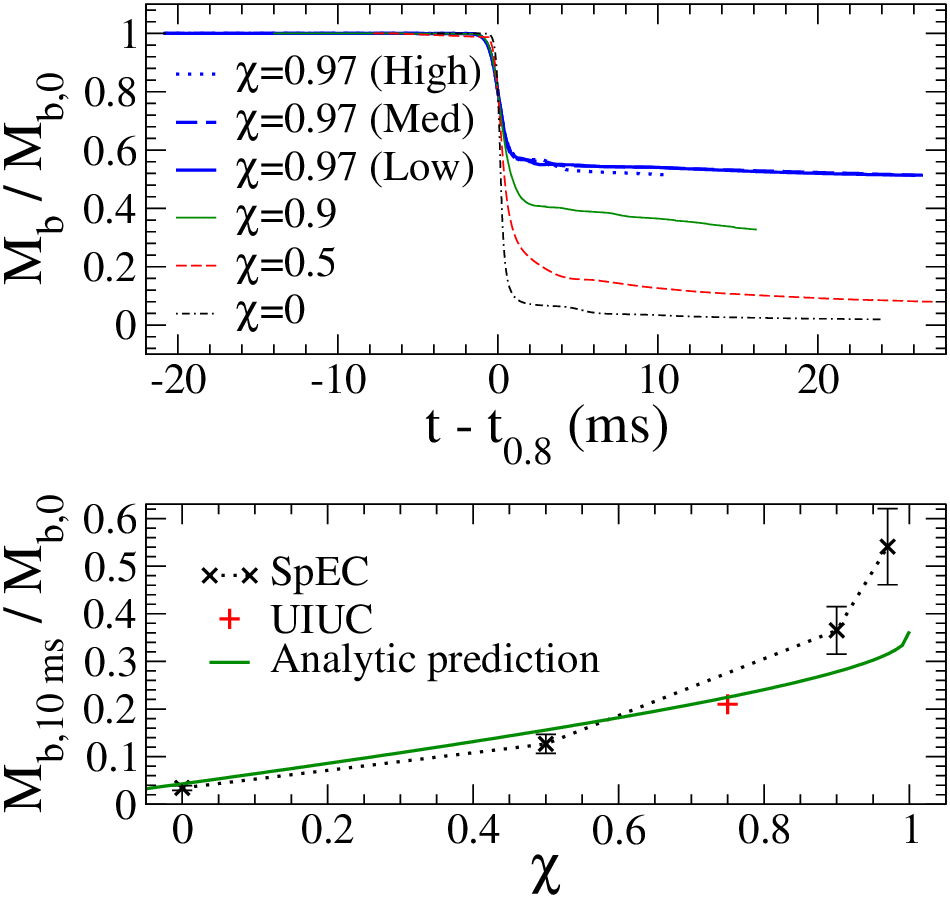}\end{center}
\caption{The evolution of the baryonic mass for black hole--neutron star 
mergers with different initial black-hole spins of magnitude $\chi$ 
aligned with the orbital angular momentum. 
The mass for $\chi=0.97$ is shown for the three different resolutions 
(low, medium, high) simulated. The masses from mergers with 
$\chi=0$, $\chi=0.5$, and $\chi=0.9$ 
are taken from Fig.~5 of Ref.~\cite{Foucart:2010eq}. 
\emph{Top panel:} 
The baryonic masses 
are shown as a function of time $t-t_{0.8}$, where $t_{0.8}$ is the time at 
which the baryon mass outside the hole 
has fallen to 80\% of its initial value. 
\emph{Bottom panel:} The baryonic mass at time $t=t_{0.8}+10\mbox{ ms}$ 
(black crosses) as 
a function of the initial black-hole spin. The medium resolution of the 
$\chi=0.97$ simulation is shown. Relative errors of $15\%$ are used
for all simulations, an estimate obtained by extrapolating
our results to infinite resolution while assuming second order
convergence of the code.
For comparison, 
also shown are i) the final disk mass obtained by the UIUC group in 
Fig.~13 of Ref.~\cite{Etienne:2008re} for case ``B'' and 
ii) the theoretical prediction derived in Ref.~\cite{Foucart2012}. 
\label{fig:BaryonMass80} 
}
\end{figure}

We should note that the amount of matter remaining outside of the
black hole after merger is not only larger than in any previous
studies of BHNS mergers: it is also significantly above the
predictions obtained by extrapolating results for lower spin black
holes in the high spin regime.  The remnant mass for lower spin
systems is generally well approximated by~\cite{Foucart2012} 
\begin{eqnarray}
\frac{M_{\rm rem}}{M_{\rm NS}} & = & 0.288 \left(\frac{3M_{\rm BH}}{M_{\rm
    NS}}\right)^{1/3} \left(1-2\frac{M_{\rm NS}}{R_{\rm NS}}\right) 
- 0.148 \frac{R_{\rm ISCO}}{R_{\rm NS}},
\end{eqnarray}
with $R_{\rm ISCO}$ the
radius of the innermost stable circular orbit around the black
hole. Figure~\ref{fig:BaryonMass80} 
shows the predicted remnant mass for
BHNS mergers at $q=3$, $M_{\rm NS}/R_{\rm NS}=0.144$, together with
numerical results at $\chi =
(0,0.5,0.75,0.9,0.97)$~\cite{Etienne:2008re,FoucartEtAl:2011}. We see
that the analytical predictions perform well for low-spin black holes
($\chi<0.9$), while strong deviations are visible for the
higher spin configurations. This is not surprising: the analytical
results are only supposed to be valid for relatively low mass remnants
($M_{\rm rem}<0.2M_\odot$).  But it does emphasize the need for
further studies of BHNS mergers at high black-hole spin in order to
properly model the characteristics of the post-merger remnant in that
regime --- especially considering that the most energetic SGRBs
observed might require the presence of such high-mass accretion
disks~\cite{Giacomazzo2012}.

\subsection{Ejecta}\label{sec:ejecta}

In addition to the formation of an accretion disk, the tidal
disruption of the neutron star can lead to the ejection of a
significant amount of unbound material. This material is initially
neutron rich, and its radioactive decay can lead to detectable optical
afterglows (``kilonovae''~\cite{Roberts2011,metzger:11}), as
well as contribute to the formation of heavy elements (r-process
nucleosynthesis). The deceleration of the ejecta in the interstellar
medium can also cause emission in the radio
band~\cite{metzger:11}. For binary neutron 
stars~\cite{hotokezaka:13}
and low-spin BHNS~\cite{Duez:2009yy} binaries, general relativistic
simulations show that only a small fraction of the neutron star
material ($<1\%$) is unbound, with kinetic energies of $10^{49}\,{\rm
  ergs}-10^{51}\,{\rm ergs}$~\cite{hotokezaka:13}.

Measurements of the unbound mass are fairly inaccurate in our code, 
especially for
very energetic and massive ejecta. This is due to the fact that the
fluid equations are only evolved up to $20M_{\rm BH}-30M_{\rm BH}$
from the center of mass of the system, while measurements of the
properties of the ejecta are best done far away from the black hole
and in low-density tidal tails (the condition $u_t<-1$, used to
determine whether material is unbound, is only valid for a stationary
metric and pressureless fluid). Additionally, the accuracy of the
evolution far away from the black hole is lower than in the forming
disk, where most grid points are concentrated. The development of
techniques allowing us to efficiently evolve the fluid equations at
larger distances without losing accuracy close to the black hole
(e.g. adaptive mesh refinement) will be required to avoid these
issues. In a recent 
%paper~\cite{2012arXiv1212.4810F}, 
paper~\cite{Foucart:2013a}, 
we showed that
the main source of error when measuring the mass of the ejecta is
generally the large grid spacing in the far zone.  The same holds in
this simulation: we find \beq M_{\rm ej}=0.26M_\odot \pm 0.16M_\odot
\eeq assuming 2nd order convergence between the medium and high
resolution.  This confirms that BHNS mergers with rapidly rotating
black holes are significantly more favorable to the ejection of
neutron-rich material. In~\cite{Foucart:2013a}, we found that
for higher mass ratio systems ($q=7$) and lower spins $\chi=0.9$, 
a few percent of a solar mass is likely to be unbound,
already a noticeable improvement compared to low-spin systems.  The
ejecta in the case studied here, which is of course an extremely
favorable configuration, is an order of magnitude larger --- thus
showing that for rapidly spinning systems, it is conceivable that a
large fraction of the mass remaining outside of the black hole at late
time is unbound.

Measurements of the kinetic energy of the ejecta are even less
reliable than its mass.  Indeed, the kinetic energy is dominated by
the most relativistic parts of the ejecta, which are also the most
poorly resolved. From measurements of $u_t$ as the outward-moving material 
is leaving the finite-difference grid, 
we estimate that the ejected material has a median
velocity $v\sim 0.5c$, and kinetic energy $E_{\rm ej}\sim
10^{52}\,{\rm ergs}-10^{53}\,{\rm ergs}$. These numbers should, however, be
considered only as order of magnitude estimates. Nonetheless, 
this is an extremely large amount of energy, which would cause the emission
of a detectable radio signal for a large fraction of the mergers within the range of Advanced LIGO as the ejecta slows down in the interstellar medium.

\section{Conclusion}
\label{sec:conclusion}

We have simulated the merger of a black hole-neutron star system with
a premerger black-hole spin parameter of $\chi=$0.97, the highest spin
yet attempted for modeling such systems.  Given the strongly nonlinear
dependence of many aspects of the merger on $\chi$ (black-hole
rotation energy, ISCO location, accretion efficiency), this simulation
carries the numerical exploration of BHNS mergers into a distinctly
new regime not sampled even by our previous simulations with $\chi=$0.9.
These nonlinearities manifest themselves clearly in our results.

Upon disruption of the neutron star, less than half of the nuclear
matter is promptly accreted into the black hole.  Almost 20\% appears
to be ejected from the system in an unbounded tidal tail outflow, and the
rest settles into a massive accretion disk.  Both the ejecta and disk
masses are higher than any previous fully relativistic BHNS simulation.  This
in itself is unsurprising; both theoretical considerations and
previous simulations would lead us to expect that in higher-$\chi$
systems more matter should evade prompt accretion.  What is notable is
that the disk mass exceeds even the expectations for this spin based
on extrapolating the trends of lower-spin systems.  If such systems
occur in nature, they must be particularly spectacular events even by
the standard of BHNS mergers.

We have closely followed the behavior of the horizon and 
of the accretion disk. 
Both merger and accretion have the net effect of decreasing 
the dimensionless spin $\chi$ of the black hole.
While the disk is thick and self-gravitating, it
appears to be quite stable, settling to an axisymmetric
quasistationary state and evolving only slowly under the influences of
outer-disk settling and accretion-induced mass decrease.  Its
subsequent evolution will be driven in part by physical processes not
included in these simulations.

That existing numerical relativity techniques can successfully treat
such an extreme system essentially unaltered is encouraging.  Further
studies of the extreme-spin regime of BHNS parameter space can now be
attempted; particularly interesting will be the exploration of
systems with higher black-hole mass.  These are expected to be more
common astrophysically, and they may also be more intriguing from a
relativist's point of view, since the disruption event happens nearer
to the horizon, as measured by the disruption separation divided by
$M_{\rm BH}$.  The investigation of the accretion system studied in
this paper should also be completed by incorporating the remaining
crucial physics, particularly the magnetic turbulence.

% \begin{acknowledgments}
\ack We are pleased to thank Christian Ott, Robert Owen, 
and Saul Teukolsky for helpful discussions. 
This work was supported in part by grants
from the Sherman Fairchild Foundation to Cornell and Caltech, 
by NSF Grants No. PHY-0969111 and
No. PHY-1005426 at Cornell; by NSF Grants No.
PHY-1068881 and PHY-1005655 at Caltech;
by NASA Grant No. NNX09AF96G at
Cornell; and NASA Grant No. NNX11AC37G and NSF Grant PHY-1068243 to WSU. 
The numerical computations presented in this paper were
performed primarily on the Caltech compute cluster \textsc{zwicky},
which was funded by the Sherman Fairchild Foundation and the NSF
MRI-R\textsuperscript{2} grant No. PHY-0960291 to Caltech.
% \end{acknowledgments}
\section*{References}
\bibliographystyle{unsrt} \bibliography{References/References}
\end{document}